\documentclass{an}
\usepackage{graphicx}
\usepackage{times}
\usepackage{fancyhdr}

\usepackage[figuresright]{rotating} 
\usepackage{natbib}
\bibpunct{(}{)}{;}{a}{}{,}

\newcommand{\aap}{A\&A }
\newcommand{\aaps}{A\&AS }
\newcommand{\azh}{AZh }

\title{Applicability of colour index calibrations to T Tauri stars}
\author{Torsten Sch\"oning\inst{1} \and Matthias Ammler\inst{1,2}} 
\institute{Astrophysical Institute and University Observatory Jena, Schillerg\"a{\ss}chen 2-3, 07745 Jena, Germany
\and Centro de Astronomia e Astrof{\'i}sica da Universidade de Lisboa, Tapada da Ajuda, 1349-018 Lisboa, Portugal}
\date{Received $<$date$>$; accepted $<$date$>$; published online $<$date$>$}
\correspondence{ammler@oal.ul.pt}
\abstract{We examine the applicability of effective temperature scales of several broad band colours to T Tauri stars (TTS). We take into account different colour systems as well as stellar parameters like metallicity and surface gravity which influence the conversion from colour indices or spectral type to effective temperature. \\
For a large sample of TTS, we derive temperatures from broad band colour indices and check if they are consistent in a statistical sense with temperatures inferred from spectral types. There are some scales (for $V-H$, $V-K$, $I-J$, $J-H$, and $J-K$) which indeed predict the same temperatures as the spectral types and therefore can be at least used to confirm effective temperatures.\\
Furthermore, we examine whether TTS with dynamically derived masses can be used for a test of evolutionary models and effective temperature calibrations. We compare the observed parameters of the eclipsing T Tauri binary V1642 Ori A to the predictions of evolutionary models in both the H-R and the Kiel diagram using temperatures derived with several colour index scales. We check whether the evolutionary models and the colour index scales are consistent with coevality and the dynamical masses of the binary components. It turns out that the Kiel diagram offers a stricter test than the H-R diagram. Only the evolutionary models of \cite{BCAH98} with mixing length parameter $\alpha=1.9$ and of \cite{DM94,DM97} show consistent results in the Kiel diagram in combination with some conversion scales of \cite{HBS00} and of \cite{KH95}.
}
\begin{document}
\maketitle
\section{Introduction}

Only in a few cases the effective temperature of T Tauri stars (TTS) can be measured directly. In most other cases, conversions from spectral type to effective temperature are used \citep[see for example][]{CK79,HSS94}. However, the spectral type of TTS is not always well determined due to their variability. For example, the classical TTS \object{DL~Tau} is classified as GVe\,--\,K7Ve \citep{CGCVS}. Thus, we recommend to verify the spectral type temperature with at least one colour index conversion scale. In this paper we want to carefully examine which colour indices are most suited on variable young stars with possible UV and IR excesses. \par
To illustrate the effects of the variability of TTS, two examples shall be given. The classical TTS (cTTS) most variable in $V$, \object{DR~Tau}, has a variability in $(B-V)_\mathrm{J}$ in the range $-0.76\leq(B-V)_\mathrm{J}\leq1.11$ \citep[for details, see][]{variability}. This difference corresponds roughly to the difference between an O-type and a K-type star. As expected, the variability of weak-line TTS (wTTS) is much smaller. The most variable wTTS, \object{V410~Tau}, has a variability of $0.87\leq(B-V)_\mathrm{J}\leq1.32$ \citep[][]{variability} which corresponds roughly to the difference between an early K-type and a late K-type star. In contrast, the typical measurement error of a colour index is only about 0.02 magnitudes.\par
The applied colour index scales are summarised in Sect. \ref{scales}; the way we calculated the colour index temperatures is explained in Sect. \ref{Tfi}.\par
In Sect. \ref{test_single}, a statistical test for a sample of apparently single TTS is done by comparing their colour index temperatures with their spectral type temperatures. We have not found any similar test in the literature.\par
If, in addition to the temperature, another stellar parameter like $\log{g}$ or $L_\mathrm{bol}$ is known, one can estimate the mass and the age of a star by means of  evolutionary models. If furthermore a mass determination independent from evolutionary models is known, one can check whether predicted and measured mass are consistent. For binaries, one can furthermore check whether both components are coeval. 
By using various conversion scales, we test combinations of evolutionary model and conversion scale (for details, see Sect. \ref{test_evolve}).
\section{Description of the applied scales} \label{scales}
 For our examination we used a compilation of common colour scales. The aim of our work is not a complete study of colour scales -- instead we are only interested which scales are best suited to infer the
temperature of TTS. We selected only scales for broad band colour indices covering (at least) most of the spectral range of TTS. For example, the scale of \cite{KH95} for $V-N$ is excluded because it only covers the range from A0 to G1. The main properties of these scales are summarised in Table \ref{table}. In this table we introduce abbreviations for the scales containing the abbreviated references and the used colour index. The last letter of these abbreviations indicates whether it is a giant (``g''), intermediate (``i''), or dwarf (``d'') scale. In case of ambiguity due to the use of multiple scales the abbreviations are extended by an expression in brackets giving either the number or the spectral range of the corresponding scale, e.\,g. AAMR99g(8).\par
In the literature, the conversion scales are given either as polynomial scales or as tabulated scales. We did not only adopt scales from the literature but in addition we defined combined scales for such colour indices which are used for at least two scales in order to be able to compare the different colour indices directly.
Annotations to the different types of scales are given in the following subsections.
\subsection{Polynomial scales}
The scales of \cite{DiB98}, \cite{BLG94}, \cite{GCC96}, and \cite{HBS00} were derived by fitting one or more polynomials $T_\mathrm{CI}(X-Y)$\footnote{In this paper, $X$ refers to any broad band magnitude and $X-Y$ refers to any broad band colour index} only to measured colour indices and temperatures of individual stars. \par
\cite{AAMR96,AAMR99} and \cite{SF00} use the metallicity as additional parameter for the polynomial in order to minimise scatter. \cite{SF00} is the only scale using also the surface gravity $\log{g}$ as additional parameter. Thus our own work has to account for the metallicity and the surface gravity in order to infer the effective temperature of a certain TTS. As the metallicity of TTS is usually not known, we take the extreme values for stars of the thin disc as measured by \cite{Fuhrmann04} as an estimate ($\mathrm{[Fe/H]}=-0.05 \pm 0.55$).  If no measurements for the surface gravity are available, i.\,e. for non-eclipsing binaries and single stars, $\log{g}=3\pm2$ is taken as rough estimate containing all values predicted by evolutionary models for TTS. For the scale of \cite{SF00}, this quite large error yields a maximum error in the temperature of only 170\,$K$ so that we can apply indifferently either giant and dwarf scales to TTS. \par
The temperatures {which are used to calibrate the scales} are usually derived by means of the InfraRed Flux Method \citep[IRFM, ][]{AAMR99-T}, scaled to direct measurements of the temperature. It is noteworthy that \cite{GCC96} lowered the IRFM temperatures of \cite{BG89} by 122\,K to adjust them to the temperatures of \cite{BLG94} while \cite{BG89} themselves suggest a lowering by only 80\,K.\par
The photometric data used to derive these scales was taken from own measurements or from the literature. Thereby the literature values were often converted from other broad band filter systems. The $K$ magnitudes of \cite{BLG94} were converted from the narrow band $Kn$ of \cite{Selby88}; the $B$ magnitudes used by \cite{SF00} were presumably converted from Str\"omgren photometry of \cite{HM98}. \cite{GCC96} do not mention the sources of their photometry.\par
The scales are based on different stellar samples. Between 13 and more than 500 stars were used. \par
Usually only a constant relative or absolute intrinsic error for the whole relation is given. On the other hand, \cite{GCC96} and \cite{HBS00} give errors for the coefficients of the polynomial. This yields larger errors (sometimes more than 1000\,K) for cooler stars.
\subsection{Tabulated scales}
We consider two tabulated scales \citep{HSS94,KH95} giving temperatures, intrinsic colours, and bolometric corrections compiled from the literature for different spectral types.\par
The temperatures in \citet[table~A5]{KH95} were taken from \cite{SK82}. Most of the given colours were either taken directly from various sources or derived by interpolating linearly between those values. For dwarfs from F- to M-type (which we are interested in) most of the infrared colours were taken from \cite{BB88} using the Johnson-Glass filter system, supplemented by data of \cite{J66}. However, we cannot retrace how some of the given colours were derived - especially the infrared colours of late G type stars. \par
The temperatures in \citet[table~4]{HSS94} were taken from \cite{SK82} for spectral types from F0 to K4 and from \cite{B91} for K7 and integer M sub-classes. For K5, M0.5, and M1.5 the origin of the temperatures cannot be retraced. From this paper only the scales for $(R-I)_\mathrm{C}$ and $I_\mathrm{C}-J$ are used, as no other scale could be found for these colour indices. The values for $(R-I)_\mathrm{C}$ for integer M sub-classes were also taken from \cite{B91}, the origin of the other values for $(R-I)_\mathrm{C}$ as well as for all $I_\mathrm{C}-J$ cannot be retraced.\par
\cite{HSS94} as well as \cite{KH95} {do not give any} intrinsic errors for these scales. We adopt the errors derived in \cite{paper_Matthias}, namely $\Delta\log{T_\mathrm{eff}}=0.015$ for temperatures originally given by \cite{SK82} and $\Delta T_\mathrm{eff}=280\,\mathrm{K}$ for temperatures originally given by \cite{B91}.
\subsection{Combined scales}\label{combinedScales}
For these scales, the temperatures are the mean of the temperatures of all scales which yield a result for the corresponding values of $X-Y$, $\log{g}$, and [Fe/H]. The error of these combined temperatures is composed of the mean error and the standard deviation of the individual temperatures. These both errors are considered  independent of each other and thus are added quadratically. These scales are hence labelled as ``combined $X-Y$''.\par
\begin{sidewaystable*}
\caption{Principal characteristics of the scales taken from the literature. Further details see Sect. \ref{scales}. \label{table}}
{\scriptsize
\begin{tabular}{llllll}
\hline
Reference & Abbreviation(s) & Colour Indices           & Considered Range  & Type Of Conversion & Type Of Error \\
          &                 & Considered In Our Work  & Of Spectral Types &  & \\
\hline \cite{AAMR96} & AAMR96d & $(B-V)$, $(V-R)_\mathrm{J}$, $(V-I)_\mathrm{J}$,
$(V-K)$, & F0 -- K5 & polynomial & standard deviation (in Kelvin) \\
 & & $(R-I)_\mathrm{J}$, $(J-H)_\mathrm{TCS}$, $(J-K)_\mathrm{TCS}$ & & \vspace{0.05cm}$ {\theta}
= \frac{5040}{T_\mathrm{eff}} = \mathrm{f}(X-Y,\mathrm{[Fe/H]}) $ & for whole relation\\
 \cite{AAMR99} & AAMR99g & $(U-V)_\mathrm{J}$, $(B-V)_\mathrm{J}$, $(V-R)_\mathrm{J}$,
$(V-I)_\mathrm{J}$, & F0 -- K5 & polynomial & standard deviation (in Kelvin) \\
 & & $(V-K)_\mathrm{TCS}$, $(V-L')_\mathrm{TCS}$, $(R-I)_\mathrm{J}$, $(I-K)_\mathrm{J}$, & & $
{\theta} = \frac{5040}{T_\mathrm{eff}} = \mathrm{f}(X-Y,\mathrm{[Fe/H]}) $ & for whole relation\\
 & & $(J-H)_\mathrm{TCS}$, $(J-K)_\mathrm{TCS}$, & & & \\
\cite{DiB98} & Ben98d & $(V-K)_\mathrm{J}$ & F -- K & polynomial & error in percent \\
             & Ben98i & & & $\log{T_\mathrm{eff}}= \mathrm{f}(X-Y)$ & for whole relation \\
             & Ben98g & & & & \\
\cite{BLG94} & BLG94i & $(V-K)_\mathrm{J}$ & A -- M & polynomial & standard deviation (in percent)\\
& & & & ${T_\mathrm{eff}}= \mathrm{f}(X-Y)$ & for whole relation \\
\cite{GCC96} & GCC96d & $(B-V)_\mathrm{J}$, $(V-R)_\mathrm{J}$, $(R-I)_\mathrm{J}$,
$(J-K)_\mathrm{J}$, & F -- K & polynomial & standard deviation of the \\
& GCC96g & $(V-K)_\mathrm{J}$ & & ${T_\mathrm{eff}}= \mathrm{f}(X-Y)$ & coefficients of the polynomial \\
\cite{HSS94} & HSS94d & $(R-I)_\mathrm{C}$, $I_\mathrm{C}-J$ & F0 -- M6 & tabulated &
$\Delta\log{T_\mathrm{eff}}=0.015$ for F0 -- K5$^1$ \\
 & & & & &$\Delta T_\mathrm{eff}=280\,\mathrm{K}$ for K7 -- M6$^1$ \\
\cite{HBS00} & HBS00d & $(U-V)_\mathrm{J}$, $(B-V)_\mathrm{J}$, $(V-R)_\mathrm{C}$,
$(V-I)_\mathrm{C}$, & F -- K & polynomial & standard deviation of the \\
& HBS00g & $(V-K)_\mathrm{CIT}$, $(V-K)_\mathrm{JG}$, $(J-K)_\mathrm{CIT}$, $(J-K)_\mathrm{JG}$ &
& ${T_\mathrm{eff}}= \mathrm{f}(X-Y)$ & coefficients of the polynomial \\
\cite{KH95} & KH95d & $(U-V)$, $(B-V)$, $(V-R)_\mathrm{C}$,
$(V-R)_\mathrm{J}$, & F0 -- M6 & tabulated & $\Delta\log{T_\mathrm{eff}}=0.015^1$ \\
& & $(V-I)_\mathrm{C}$, $(V-I)_\mathrm{J}$, $(V-J)_\mathrm{JG}$, $(V-H)_\mathrm{JG}$, & & & \\
& & $(V-K)_\mathrm{JG}$, $(V-L)_\mathrm{JG}$, $(V-M)_\mathrm{JG}$ & & & \\
\cite{SF00} & SF00i & $(B-V)$ & F0 -- K5 & polynomial & rms of residuals \\
 & & & & $\log{T_\mathrm{eff}}= \mathrm{f}(X-Y,\mathrm{Fe/H},\log{g})$ & \\
\hline
\end{tabular}}
\begin{flushleft}
$^1$ -- Errors are taken from \cite{paper_Matthias}
\end{flushleft}
\end{sidewaystable*} 

\section{{Remarks on the} calculation of the colour index temperatures $T_\mathrm{CI}$ }  \label{Tfi} 
In an ideal case, the colour index conversion scales should be only used for observations done with the same filter system. This is in general not possible because there are too many different filter systems. A consequent conversion into only {\em one} standard filter system is not possible, because often either: (a) The observational data are given in instrumental filter systems for which a conversion formula is not known. (b) The photometry was already converted from another filter system. With a further conversion one would have twice the statistical error of such a conversion. Or (c) the used filter system is even not given.\par
Detailed information about the different filter systems can be found in \cite{MM00} and \cite{ADPSII}\footnote{The filter system ``\cite{BB88}'' given in \cite{MM00} and \cite{ADPSII} is equal to the Johnson-Glass filter system.}; information about the TCS filter system can be found in \cite{AAMR94,AAMR98}.\par
Especially for the R and I band the differences between the distinct filter systems are not negligible. Considering $V-R=0.50$, for example, the difference between the Cousins and {the} Johnson filter system yields a difference of about 1000\,K in the resulting colour index temperature. Thus, in this paper, (RI)$_\mathrm{C}$\footnote{For most TTS, $R$ and $I$ have been measured in the Cousins system, probably due to the larger transmission of these filters.} and (RI)$_\mathrm{J}$ are treated as completely different passbands. Observations in other RI filter systems are ignored because corresponding colour index scales do not exist. All other passbands are always treated as if they were Johnson, {adding a negligible error}.\par
The given colour indices have to be corrected for interstellar reddening, i.\,e. they have to be converted to intrinsic colours. Therefore, one has to calculate the (filter system dependent) extinction $A_X$ from the visual extinction $A_V$ which is usually given. For this purpose, the $\frac{A_X}{A_V}$ of \cite{ADPSII} for an interstellar reddening $R_V = 3.1$ are used. Possible systematic errors in the determination of the $A_V$ cannot be considered in this paper.
\section{The statistical test with single TTS} \label{test_single}
For a sample of about 150 apparently single TTS we calculated the difference $D$ between the colour index temperature $T_\mathrm{CI}$ and the mean spectral type temperature $\overline{T_\mathrm{S}}$ for each star and for each colour index conversion scale.
\begin{equation}
D=T_\mathrm{CI}-\overline{T_\mathrm{S}}
\end{equation}
$\overline{T_\mathrm{S}}$ is derived with the conversion scales of \cite{B79,B91,CK79,SK82,dJN87,HSS94,KH95,L99}; and \cite{P98}. Thereby, the given spectral types are converted into spectral numbers \citep[following][]{dJN87}. For example, M1 is converted to $6.7$. If necessary, we interpolated linearly, with spectral types of the conversion scales being converted into spectral numbers. If intervals are given, the mean spectral number is used. For example, instead of ``K7-M0'' we use the spectral number $6.35$, which corresponds to ``K8.5''. The error of the spectral number of our sample stars is either assumed to be 0.05 (according to roughly half a subclass) or as half the given interval. The $\overline{T_\mathrm{S}}$ uncertainty $\Delta\overline{T_\mathrm{S}}$ is composed of the mean intrinsic error of the used scales \citep[see][table 2, for details]{paper_Matthias} and the standard deviation of the individual spectral type temperatures.\par
We then used four criteria to test whether a scale is applicable to TTS or not (see Sect. \ref{Kriterien} for details).\par
For comparison, we did the same test with a sample of about 250 non-variable main sequence and / or giant stars (hence referred to as ``old stars''). Of course, in this case dwarf scales were only used for dwarfs and giant scales only for giants.\par
The data for our sample of single TTS (photometry, $A_V$, and spectral type) were taken from a compilation of Ralph Neuh{\"a}user (priv. comm.)\footnote{which comprises data from \cite{Basri94}, \cite{BBM92}, \cite{Bric93,Bric98,Bric99}, \cite{CK79}, \cite{Elias78}, \cite{Giziz99}, \cite{Gom92}, \cite{Haas90}, \cite{pat93}, \cite{HSS94}, \cite{HSKJ91}, \cite{HBC}, \cite{HVR86}, \cite{JH79}, \cite{KHSS90,Kenyon93,Kenyon94b,Kenyon94}, \cite{LeiHas89}, \cite{L91,L93}, \cite{MaMa94}, \cite{Mar94}, \cite{MoZ91}, \cite{My87}, \cite{RdHw199}, \cite{Rydgr84}, \cite{Si92,Simon95}, \cite{Skr90}, \cite{SS89}, \cite{Str90}, \cite{SS93}, \cite{Torr95}, \cite{Vrba85}, and \cite{W88}.} We have taken the data for the old stars from the ``Catalogue of the Brightest Stars'' of \cite{Ochsenbein88}, whereas only non-variable stars with known $V$ magnitude, spectral type F0 or later, and luminosity class III to V were used. The (UBVRI)$_\mathrm{J}$ photometry of this catalogue was supplemented with 2MASS JHK photometry. The visual extinction $A_V$ of the old stars was calculated from the Hipparcos parallax and the distance dependent extinction $\gamma=0.3\ \mathrm{mag}\,\mathrm{kpc}^{-1}$ (valid in the galactic plane outside of dark clouds). For both samples, 0.02 magnitudes are adopted as typical measurement error of the resulting intrinsic colours.
\subsection{Criteria for an applicable scale} \label{Kriterien}
\begin{figure}
\resizebox{\hsize}{!}{
\includegraphics[angle=90, width=17cm, clip]{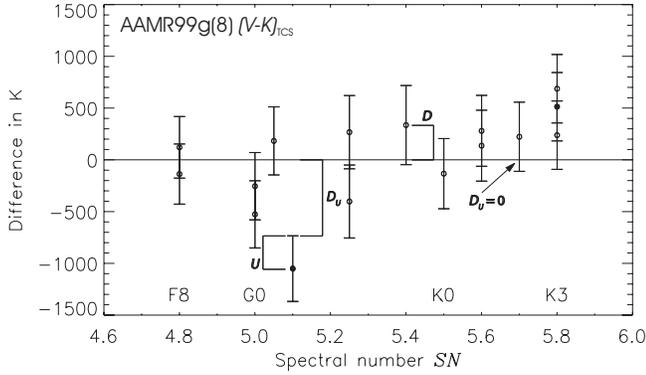} } \caption{Example for a 
plot of the differences $D$ and $D_U$ as well as the uncertainties $U$\ over spectral number $S\!N$ for the scale \#8 of \cite{AAMR99} - AAMR99g(8) - for the colour index $(V-K)_\mathrm{TCS}$. wTTS are marked with open circles, cTTS are marked with full circles. This scale was found to fulfil our criteria, i.\,e. to be applicable to TTS.} \label{D_DU}
\end{figure}

In order to quantify discrepancies between $T_\mathrm{C}$ and $\overline T_\mathrm{S}$, the uncertainty $U$ of the differences $D$ is calculated from the quadratically added errors of both temperatures:
\begin{equation}
U= \sqrt{\Delta^2T_\mathrm{CI}+\Delta^2\overline{T_\mathrm{S}}} \ .
\end{equation}
{ We account for the heterogeneous errors in the test and in principle we regard the colour index temperature consistent with the spectral type temperature if:
\begin{equation}
|D|<U
\end{equation}
In order to perform statistical tests, a difference $D_U$ is now defined which is the remaining difference when the uncertainty $U$ has already been subtracted from the absolute value of $D$:}
\begin{eqnarray}
D_U \equiv& D-U&\mathrm{,if\ } U \leq |D| \mathrm{\ and\ } D > 0 \nonumber \\
D_U \equiv& D+U&\mathrm{,if\ } U \leq |D| \mathrm{\ and\ } D < 0 \nonumber \\
D_U \equiv& 0  &\mathrm{,if\ } U > |D| \ \ \ \ .
\end{eqnarray}
For each scale, the mean $\overline{D_U}$ and the standard deviation $\sigma(D_U)$ is calculated. Furthermore, the differences $D$ and their uncertainties $U$ are plotted versus spectral number $S\!N$ {in order} to calculate the slope of $D(S\!N)$ {and} to recognise possible systematic trends. An example of such a plot is given in Fig. \ref{D_DU}.\par
Finally, we used the following four criteria to decide whether a scale is applicable to TTS and old stars, respectively.
\begin{enumerate}
\item The mean difference $\overline{D_U}$ should not be significantly different from zero. $\overline{D_U} = 0$ means that $\overline{T_\mathrm{S}}$ and $T_\mathrm{CI}$\ are equal in average as is expected if both represent the effective temperature. A t-test with $\alpha=0.01$ and $\alpha=0.05$, respectively, is done to check for significant differences. 
\item According to \cite{Moore}, a t-test for less than 15 data points is only valid if the data is normally distributed. As this cannot be assumed a priori for our sample, only scales yielding results for at least 15 stars are taken into account.
\item $D(S\!N)$ should have no significant slope, i.\,e. the slope should be less than 3\,$\sigma$ different from constancy { where $\sigma$ is the uncertainty estimate of the slope}. 
\item The mean error of the colour index scale $\overline{\Delta T_\mathrm{CI}}$ should be less than 500\,K. This is rather a practical question than a statistical criterion.
\end{enumerate}
\subsection{Results}
\subsubsection{Results for old stars}
\begin{figure*}
\includegraphics[height=21.8 cm, clip] {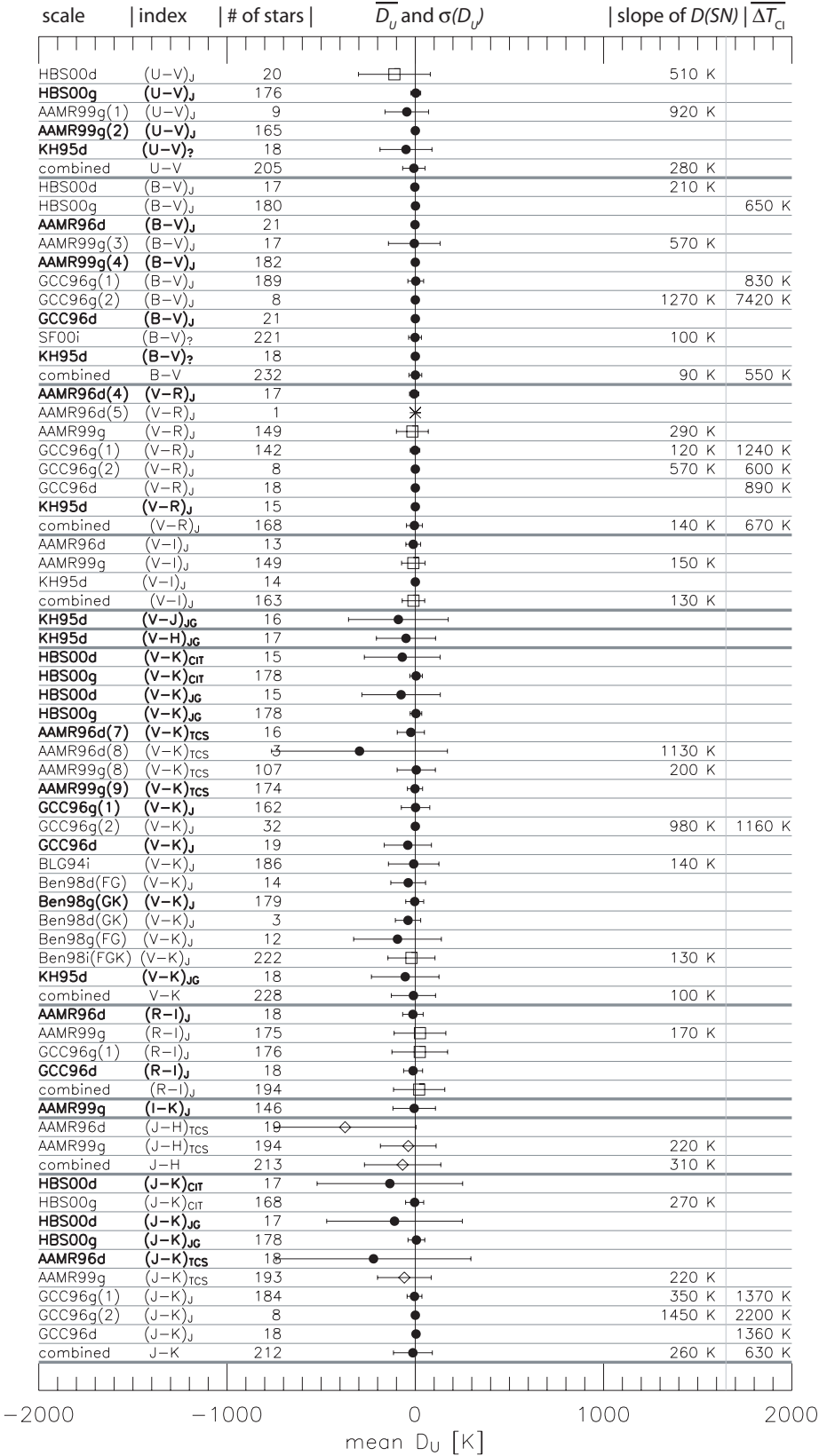} 
\caption{The results of the statistical test for old stars. The first column gives the abbreviation of the used scale and the used colour index. Scales which pass our test are marked bold. The second column gives the number of stars the scale yields a result for. The third column gives a graphical representation of the $\overline{D_U}$ (symbols) and of the $\sigma(D_U)$ (error bars). Filled circles mean that there is no significant difference. Open symbols denote a significant difference with a probability of error $\alpha=0.01$ (diamonds) and $\alpha=0.05$ (squares), respectively. The fourth column gives the slope of the $D(S\!N)$ if it is significant and the fifth column the mean error $\overline{\Delta T_\mathrm{CI}}$ of the colour index scale, if larger than 500\,K. The unit of the slope is [K] since the abscissa is the dimensionless spectral number (see Fig. \ref{D_DU}). Further details are given in the text.} \label{results_evolved}
\end{figure*}

The detailed results are shown in Fig. \ref{results_evolved}.
As expected, spectral and colour index temperatures match quite well for the old stars because both the conversion scales for spectral types and for colour indices have been calibrated with data of old stars. However, it is noteworthy that half of the scales do not pass all four criteria although they are made for old stars.\par
Mean errors of the colour index scales larger than 500\,K appear only for the scales of \cite{GCC96} and \cite{HBS00} who give errors of the coefficients of their polynomial scales instead of errors of the whole scale. At the same time, this legitimates the quite large error of [Fe/H] ($\Delta\mathrm{[Fe/H]}=0.55$) {adopted by us} because the metallicity dependent scales of \cite{AAMR96,AAMR99} do not show too large errors.
\subsubsection{Results for TTS}
\begin{figure*}[ht] 
\includegraphics[height=21.8cm, clip] {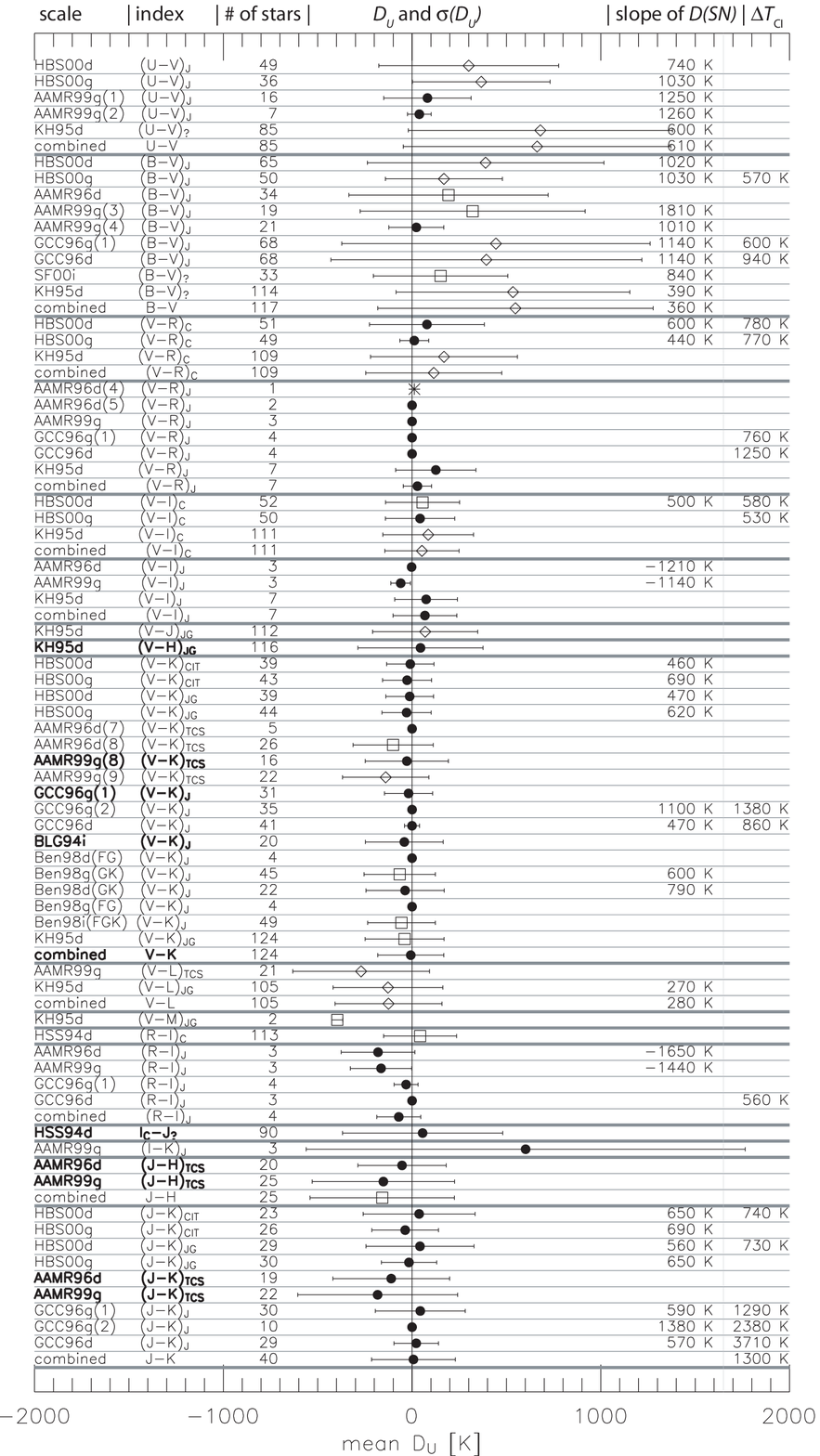} 
\caption{Same as Fig. \ref{results_evolved}, but for TTS.}
 \label{Results}
\end{figure*}

We did the test for the TTS sample as a whole as well as separately for the cTTS and wTTS sub-samples. As an example, the detailed results for one applicable scale are given in Fig. \ref{D_DU}, the summarised results for all scales and the whole TTS sample are given in Fig. \ref{Results}.
Due to the UV excess making the stars bluer, the colour index temperatures for $U-V$ and $B-V$ are in general larger than the mean spectral type temperatures. For most of the $U-V$ and $B-V$ scales, the differences $D=T_\mathrm{CI}-\overline{T_\mathrm{S}}$ increase significantly with spectral number. This corresponds to a larger UV excess (in magnitudes) for stars with weak {photospheric} UV emission (M stars) than for stars with comparatively large photospheric UV emission (earlier type stars). Analogously for $V-L$ and $V-M$, the colour index temperatures are too low due to the infrared excess.\par
For the colour indices $V-H$, $V-K$, $I_\mathrm{C}-J$, $J-H$, and $J-K$ at least some of the tested conversion scales pass all four criteria, namely KH95d $(V-H)_\mathrm{JG}$, AAMR99g(8) $(V-K)_\mathrm{TCS}$, GCC96g(1) $(V-K)_\mathrm{J}$, BLG94i $(V-K)_\mathrm{J}$, ``combined $V-K$'', HSS94d $I_\mathrm{C}-J$, AAMR96d $(J-H)_\mathrm{TCS}$, AAMR99g $(J-H)_\mathrm{TCS}$, AAMR96d $(J-K)_\mathrm{TCS}$, and AAMR99g $(J-K)_\mathrm{TCS}$. However one has to keep in mind that these criteria are only statistical ones -- individual stars sometimes have differences $D_U$ of less than $-1000$\,K or more than $+1000$\,K even if these ``applicable scales'' are used.\footnote{For the wTTS \object{[HJS91]~4423}, the spectral type used (M5) is probably wrong because {\em all} colour index temperatures are more than 1000\,K larger than the mean spectral type temperature. The observed colours are consistent on average with a spectral type of about K2 to K3 \citep[][table~A5]{KH95}.} Nevertheless, as shown above the spectral type temperature of TTS is also not always a accurate measure for effective temperature. Thus our ``applicable scales'' can be used only to verify the spectral temperature.
\begin{figure*}[ht] 
\includegraphics[height=21.8cm, clip] {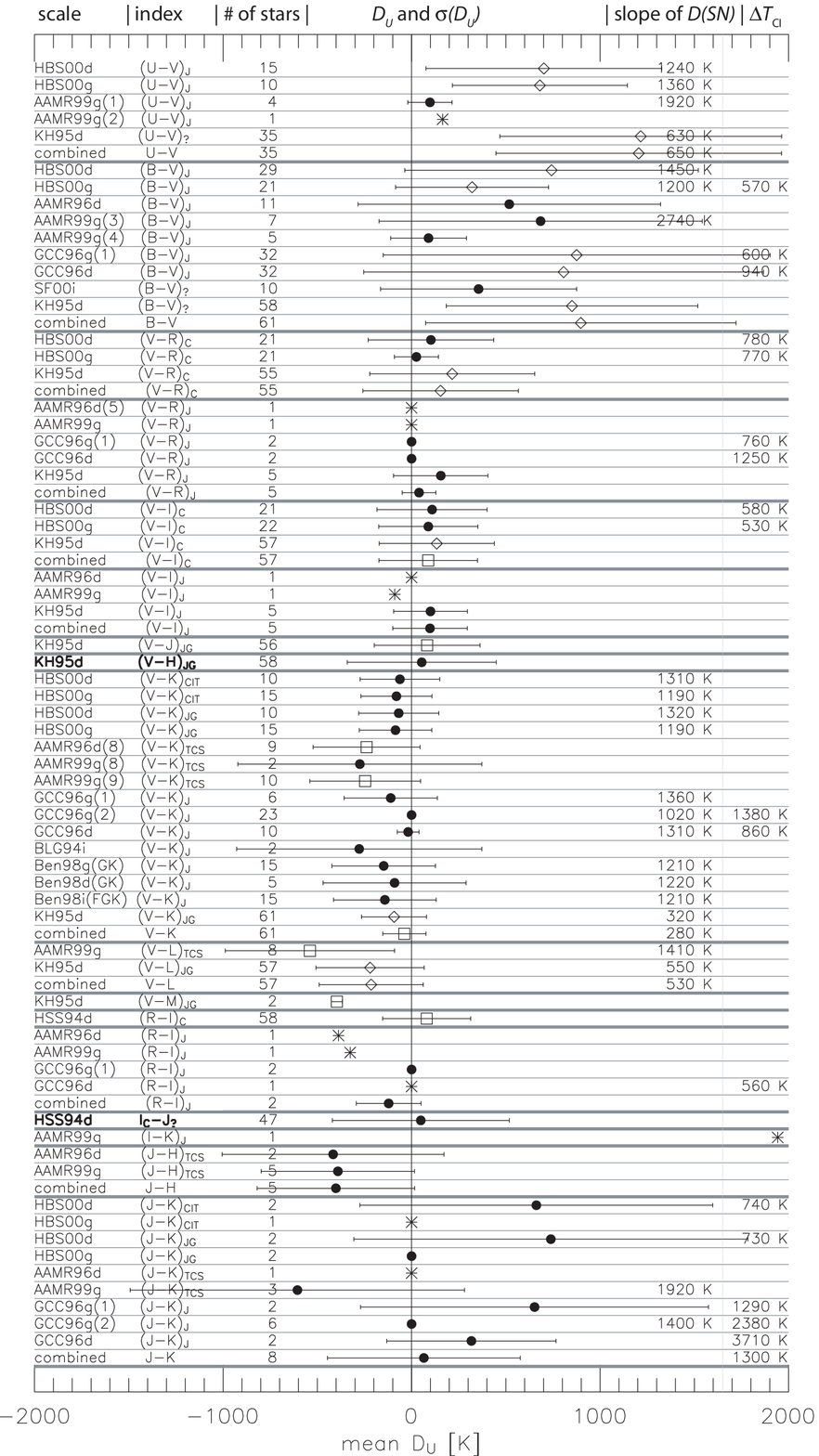}
\caption{Same as Fig. \ref{results_evolved}, but for cTTS.} \label{Results_cTTS}
\end{figure*}
\begin{figure*}[ht] 
\includegraphics[height=21.8cm, clip] {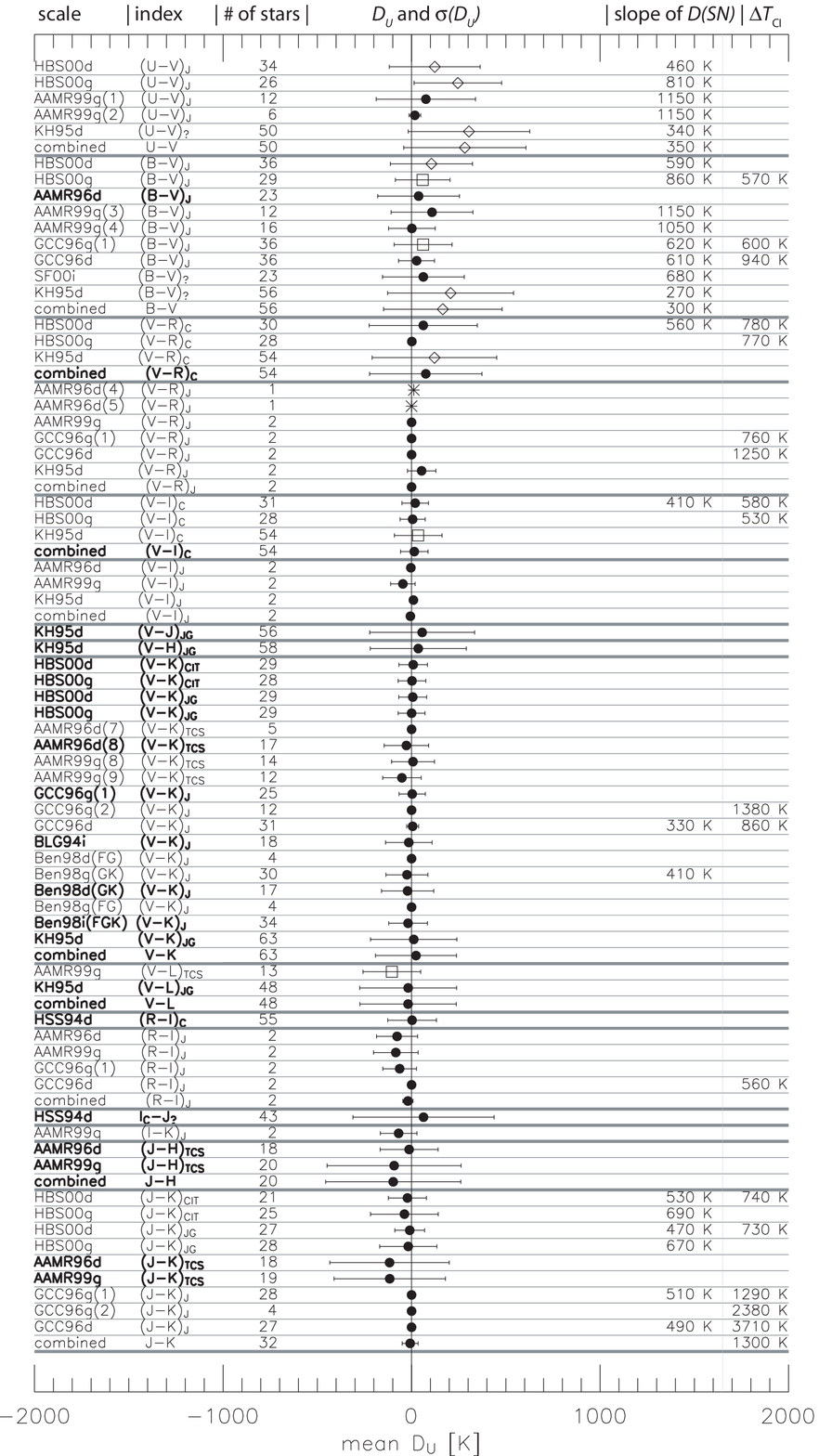} 
\caption{Same as Fig. \ref{results_evolved}, but for wTTS.} \label{Results_wTTS}
\end{figure*}

As wTTS have weaker variability and excesses than cTTS, the mean differences $\overline{D_U}$ as well as the $\sigma(D_U)$ are in general smaller for wTTS than for cTTS. Thus, if we consider only wTTS, we find 25 applicable scales { for $B-V$, $V-J$, $V-H$, $V-K$, $V-L$, $I_\mathrm{C}-J$, $J-H$, and $J-K$. For $(V-R)_\mathrm{C}$ and $(V-I)_\mathrm{C}$ only the combined scales pass our criteria. }If we consider only cTTS, we find only two applicable scales (namely KH95d $(V-H)_\mathrm{JG}$ and HSS94d $I_\mathrm{C}-J$). The results for cTTS and wTTS are shown in Fig. \ref{Results_cTTS} and Fig. \ref{Results_wTTS}, respectively.
\section{The test with evolutionary models} \label{test_evolve}
 Pre-main sequence evolutionary models give a clue to the understanding of the evolution of the very young stars. They relate age and mass of a given star with other stellar parameters, like for example the effective temperature. However, the current understanding of the pre-main sequence evolution is not sufficient in order to give predictions at the required level of accuracy. In particular, there are free parameters in the underlying physics which are not well-constrained, for example the mixing-length parameter of the description of convection. \par
In this section we want to test combinations of evolutionary model and conversion scale for TTS where mass, $T_\mathrm{CI}$, and either $\log{g}$ or $L_\mathrm{bol}$ is known.
\subsection{Suitable test objects}
In order to find suitable objects we first consider the most fundamental parameter of a star, its mass. 
According to \cite{HW04}, there are 18 TTS in 13 systems so far with known masses: six single cTTS for which the mass could be derived from the mass of the surrounding disc, three spectroscopic T Tauri binaries for which the inclination $i$ could be estimated, two eclipsing binaries with one T Tauri component and two eclipsing T Tauri binaries.\par
The single cTTS cannot be used for our test as their apparent bolometric luminosity can be determined only with large uncertainties due to the excesses and the variability of these objects. For example, \object{GM~Aur}, one of the single cTTS with known mass, has a luminosity of 1.00 $L_\odot$ \citep[according to][]{val93} or of 3.96 $L_\odot$ \citep[according to][]{HW04}. \par
The binaries can be used only for our test if resolved colours of the components are given. But this is the case for just two binaries, namely NTTS 045251+3016 and EK~Cep. Unfortunately we cannot use any of them as is explained in the following.\par
For the astrometric T Tauri binary \object{NTTS 045251+3016}, \cite{St01} give the resolved $V-H$ colours which were used to determine the temperature of the components with the conversion scale of \cite{KH95}. As we use only this scale to calculate colour index temperatures with $V-H$, we would only reproduce the results given by \cite{St01} themselves. For the eclipsing spectroscopic binary \object{EK Cep}, light curves for $B$, $V$, and $R$ were obtained by \cite{E66,EH90}; and \cite{K83a}\footnote{From the Julian dates given in table 1 of \cite{K83a} (3902.3622 to 4629.5497), probably only 2\,440\,000 was subtracted -- not 2\,444\,000 as given by \cite{K83a}. Otherwise the light curve would have been obtained from 1990 till 1991 -- too late for a paper from 1983.}. 

\cite{HE84} calculated colour indices for the secondary from the relative contributions of both components to the brightness in the B, V, and R band and the BVR magnitudes of \cite{K83a} outside and within eclipse. This object is not used as the colour indices calculated from the primary eclipse are clearly discrepant from the ones calculated from the values outside the eclipses.\par

Therefore, it seems that no suitable test object can be found. On the other hand, the program ``Nightfall'' by R. Wichmann allows calculating resolved colours from given light curves. Thus, we could derive resolved colours for the eclipsing wTTS binary \object{V1642~Ori~A} because the $BVJHK$ light curves of this object were analysed with this program by \cite{Covino04} (see Table \ref{Obj_Par}). We re-calculated their final light curve solution using the adopted values $T_1=5200$\, K, $q=0.7305$, and $e=0$ as well as the values given in \citet[table 4 {$[$}no-spots-solution{$]$} and table 5]{Covino04}. The resolved broad band magnitudes are given in Table \ref{Phot_Cov}.\par
Intrinsic colours were calculated using $E(B-V)=0.10$ as given by \cite{Covino04}. Our intrinsic colours are consistent with the colours during secondary minimum obtained by \cite{Covino04}.
\begin{table}[h]
\caption{\label{Obj_Par}The physical parameters of the components of V1642 Ori (Covino et al. 2004).}
\centering
\begin{tabular}{lll}
\hline
&Primary \object{V1642 Ori Aa}&Secondary \object{V1642 Ori Ab}\\
\hline
$M\,[M_{\odot}]$&$1.27\pm0.01$&$0.93\pm0.01$\\
$R\,[R_{\odot}]$&$1.44\pm0.05$&$1.35\pm0.05$\\
${\log}g\,[cgs]$&$4.22\pm0.02$&$4.14\pm0.02$\\
$T_\mathrm{eff}\,$[K]&$5200\pm150$&$4220\pm150$\\
spectral type&K1$\pm$1&K7-M0\\
\hline
\end{tabular}
\end{table}

\begin{table}[]
\caption{\label{Phot_Cov} Resolved broad band magnitudes of V1642~Ori~A calculated by us using the software ``Nightfall'' by R. Wichmann. Only those values are given for which light curves exist. The errors are due to the errors of the ``input data'' (mass, temperature, combined luminosity ..., see Table \ref{Obj_Par}).}
\centering
\begin{tabular}{ccc}
\hline 
         & Primary \object{V1642 Ori Aa} & Secondary \object{V1642 Ori Ab} \\
\hline $B$ & 13.74 $\pm$ 0.04    & 15.70 $\pm$ 0.16 \\
       $V$ & 12.87 $\pm$ 0.04    & 14.44 $\pm$ 0.14 \\
       $J$ & 11.19 $\pm$ 0.05    & 11.92 $\pm$ 0.10 \\
       $H$ & 10.74 $\pm$ 0.06    & 11.17 $\pm$ 0.08 \\
       $K$ & 10.60 $\pm$ 0.06    & 11.03 $\pm$ 0.08 \\ \hline
\end{tabular}
\end{table}
\subsection{The test procedure}
\begin{figure}
\resizebox{\hsize}{!}{
 \centering 
\includegraphics[width=17cm, clip]{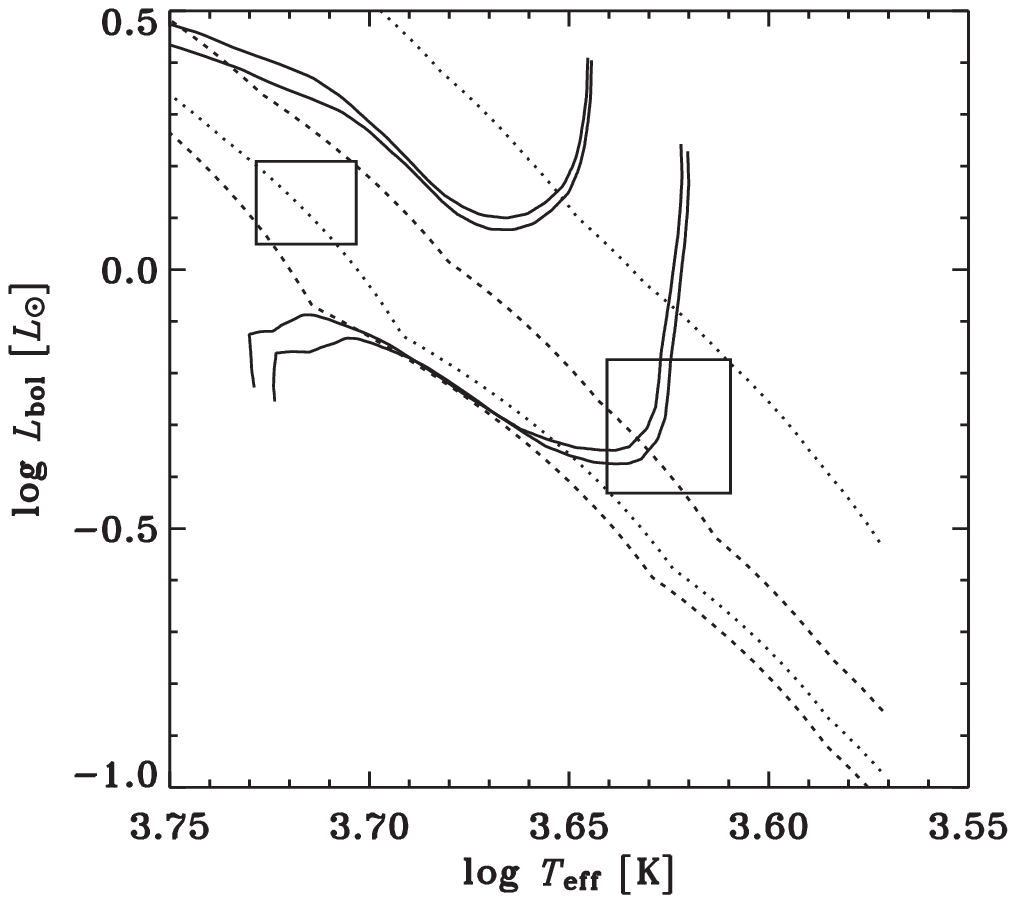} 
} 
\caption{The HR diagram with the components of V1642 Ori A and tracks and isochrones of \cite{PS99}.\newline
The error boxes represent the parameters of V1642~Ori~Aa and V1642~Ori~Ab, respectively. The effective temperatures are adopted from \cite{Covino04}.\newline 
The solid lines are tracks interpolated for the limits on the mass of these both objects. The interpolated isochrones give the minimum and maximum age for both primary (dashed line, 9 to 21 Myrs) and secondary component (dotted line, 3 to 16 Myrs).}
\label{Test_HRD}
\end{figure}
We compare the masses, radii, and luminosities of the components of V1642 Ori with the predictions of different evolutionary models in both the HR diagram and the so-called Kiel diagram (surface gravity vs. effective temperature). Thereby the comparison is not only done with the effective temperatures given by \citet{Covino04}. Instead we repeated the comparison with several sets of temperatures resulting from the application of colour index scales to the inferred colours since neither the best evolutionary models nor the best colour index scale is known. Of course this comparison does not allow one to draw conclusions on a certain scale or a certain model but only on combinations of scales and models. \par

Each component is represented by an error box in the HRD- and Kiel diagram, respectively. If (a) these rectangles intersect with the tracks for the upper and lower limit of the mass of the particular component and (b) an isochrone can be found intersecting both rectangles, then the considered combination of conversion scale and evolutionary model is assumed to be consistent with the observational data given by \cite{Covino04}. 
In order to better constrain possible masses and ages we interpolated the quite coarse grid of tracks and isochrones given by the evolutionary models. Although this refinement may be not really physical, it is more precise than just interpolating mass and age by eye. We assume that the interpolated isochrones have an error of $0.5\,$Myrs.\par

In the example shown in Fig.~\ref{Test_HRD}, we use the stellar parameters given in Table \ref{Obj_Par}. One can see that the effective temperature and luminosity of only the secondary is consistent with the values predicted by the evolutionary model. The interpolated isochrones give a coeval solution with an age of 9 to 16 Myrs.\par
The following evolutionary models were used: \cite{BCAH98}; \cite{DM94} with ``Alexander'' opacities; \cite{DM97} with $Y=0.26$ and $X_\mathrm{D}=2\cdot10^{-5}$ or $X_\mathrm{D}=4\cdot10^{-5}$, respectively, as well as with $Y=0.28$ and $X_\mathrm{D}=1\cdot10^{-5}$, $X_\mathrm{D}=2\cdot10^{-5}$,or $X_\mathrm{D}=4\cdot10^{-5}$, respectively; \cite{PS99}; \cite{SDF00}; \cite{Yi03} with $Z=0.01$, $Z=0.02$, $Z=0.04$, $Z=0.06$, and $Z=0.08$. \par
We used only those conversion scales which (a) were found to be applicable for wTTS as described above and (b) yield results for both components. This is the case for KH95d $(V-J)_\mathrm{JG}$, KH95d $(V-H)_\mathrm{JG}$, HBS00d $(V-K)_\mathrm{CIT}$, HBS00g $(V-K)_\mathrm{CIT}$, HBS00d $(V-K)_\mathrm{JG}$, HBS00g $(V-K)_\mathrm{JG}$, Ben98i(FGK) $(V-K)_\mathrm{J}$, KH95d $(V-K)_\mathrm{JG}$, and ``combined $V-K$''. 
\subsection{Results}
\subsubsection{Test with the HRD}
The comparison in the HR diagram offers no really meaningful test, mainly due to the large relative error of the luminosities. All evolutionary models give consistent results in combination with at least one conversion scale. As well, every conversion scale gives consistent results in combination with at least one evolutionary model. We get an overall range of possible ages from 2 to 51 Myrs. In combination with {\em every} conversion scale used within this test we obtain consistent results for the models of \cite{BCAH98} with $\alpha = 1.5$ and $\alpha = 1.9$, of \cite{DM94} with FST mixing theory, all models of \cite{SDF00} except for $Z=0.10$, and of \cite{Yi03} with $Z=0.04$.\par
The different evolutionary models yield different ages. The models of \cite{DM97} with $Y=0.26$ give the lowest ages, namely 3 to 8 Myrs if the temperatures given by \cite{Covino04} are used. Using the same temperatures, the models of \cite{BCAH98} with $\alpha = 1.0$ and $Y=0.275$ yield the highest ages, namely 15 to 26 Myrs.
\subsubsection{Test with the Kiel diagram}
In the Kiel diagram the conversion scales KH95d $(V-H)_\mathrm{JG}$, Ben98i(FGK) $(V-K)_\mathrm{J}$, and KH95d $(V-K)_\mathrm{JG}$ as well as the temperatures given by \cite{Covino04} yield inconsistent results for every evolutionary model. As well, the evolutionary models \cite{BCAH98} with $\alpha = 1.0$ and $\alpha = 1.5$, \cite{PS99}, \cite{SDF00}, and \cite{Yi03} yield inconsistent results for every conversion scale.
For the remaining combinations of evolutionary models and conversion scales, only few yield consistent results (see Table \ref{combinations}). The test with the Kiel diagram is more deciding because the relative errors of the surface gravities are smaller than the relative errors of the bolometric luminosities.\par
The overall range of possible ages (4 to 8 Myrs) is smaller than the range obtained with the HRD.\par
\begin{table*}[]
\caption{\label{combinations} Results of the test with the Kiel diagram for V1642~Ori~A. The {\em combinations} of evolutionary models (rows) and conversion scales (columns) yielding consistent results are marked with ``X''.\newline ``combined $V-K$'' refers to the mean temperature of all $V-K$ scales (see section \ref{combinedScales}). The parameters of the evolutionary models are as follows: $\alpha$ -- mixing length parameter, FST -- full spectrum of turbulence, MLT -- mixing length theory, $Y$ -- relative Helium abundance, $X_\mathrm{D}$ -- relative Deuterium abundance.}
\centering
{\scriptsize 
\begin{tabular}{lcccccc}
\hline
              & KH95d                & HBS00d & HBS00g  & HBS00d & HBS00g & ``combined $V-K$'' \\
              & $(V-J)_\mathrm{JG}$ & $(V-K)_\mathrm{CIT}$ & $(V-K)_\mathrm{CIT}$ & $(V-K)_\mathrm{JG}$ & $(V-K)_\mathrm{JG}$ & \\
\hline 
\cite{BCAH98} &    &    & X  &    & X & X   \\
\ \ \ \ $\alpha=1.9$  &                           &                             & &&\\
\cite{DM94}   &    &    & X  &    & X & X   \\
\ \ \ \ FST, "Alexander" opacities           &                           &                             & &&\\
\cite{DM94}   & X  & X  & X  & X  & X & X   \\
\ \ \ \ MLT, "Alexander" opacities           &                           &                             & &&\\
\cite{DM97}   &    & X  &    & X  &   & X  \\
\ \ \ \ $Y=0.26$, $X_\mathrm{D}=2\cdot10^{-5}$ &                  &&&&\\
\cite{DM97}   &    & X  &    & X  &   & X  \\
\ \ \ \ $Y=0.26$, $X_\mathrm{D}=4\cdot10^{-5}$ &                  &&&&\\
\cite{DM97}   &    & X  &    & X  &   & X  \\
\ \ \ \ $Y=0.28$, $X_\mathrm{D}=1\cdot10^{-5}$ &                  &&&&\\
\cite{DM97}   &    & X  &    & X  &   & X  \\
\ \ \ \ $Y=0.28$, $X_\mathrm{D}=2\cdot10^{-5}$ &                  &&&&\\
\cite{DM97}   &    & X  &    & X  &   & X  \\
\ \ \ \ $Y=0.28$, $X_\mathrm{D}=4\cdot10^{-5}$ &                  &&&&\\
\hline
\end{tabular} }
\end{table*} 
It is remarkable that any evolutionary model does not give consistent results if the temperatures by \cite{Covino04} are used in the Kiel diagram.
All the suitable scales except KH95d $(B-V)$ yield nearly the same primary temperature as \cite{Covino04}. On the other hand, the suitable scales imply higher secondary temperatures than \cite{Covino04} -- while all other scales do not.
\section{Conclusions}
We compiled several conversion scales which allow to derive effective temperatures from broad band colour indices, in order to examine their applicability to TTS.\par
These scales were first tested with a large sample of apparently single TTS. For this purpose we used four statistical criteria. As a result, we found ten scales for $V-H$ and $V-K$ as well as for the infrared colours $I_\mathrm{C}-J$, $J-H$, and $J-K$ which are consistent with the temperatures derived from spectral type.\par
Furthermore we compared the colour index temperatures and the dynamically derived masses of the components of the eclipsing T Tauri binary V1642~Ori~A with predictions of pre-main sequence evolutionary models \citep{BCAH98,DM94,DM97,PS99,SDF00,Yi03}, both in the HR diagram and the Kiel diagram.\par
In the HR diagram all evolutionary models give consistent results in combination with at least one conversion scale. As well, every conversion scale gives consistent results in combination with at least one evolutionary model. 
In the more decisive Kiel diagram, the evolutionary models of \cite{BCAH98} with $\alpha=1.9$, \cite{DM94}, and \cite{DM97} yield consistent results in combination with at least some conversion scales. In this diagram the scales HBS00d $(V-K)_\mathrm{CIT}$, HBS00d $(V-K)_\mathrm{JG}$, and ``combined $V-K$'' appear to be most suitable. But it is important to keep in mind that only combinations of evolutionary model and conversion scale are tested -- neither evolutionary models nor conversion scales alone. 
\par
As the Kiel diagram offers stricter constraints on evolutionary models than the H-R diagram, we recommend to use the Kiel diagram whenever possible.
\acknowledgements
We thank R. Neuh{\"a}user for giving us his compilation of data of TTS, E. Covino for providing us the light curves of V~1642~Ori~A, and R. Wichmann for his program ``Nightfall''. \par
This research has made use of the SIMBAD database, operated at CDS, Strasbourg, France and of NASA's Astrophysics Data System.\par
The comments of the anonymous referee helped to improve the manuscript substantially.
\end{document}